\documentclass[aps,prd,groupedaddress,nofootinbib,showpacs]{revtex4}
\usepackage{graphicx}

\def \wg#1{\mbox{\boldmath ${#1}$}}
\newfont{\ssg}{cmssbx10}
\def \w#1{\mbox{\ssg {#1}}}

\begin{document}


\title{Jacobi-like bar mode instability of relativistic rotating bodies}


\author{Dorota Gondek-Rosi\'nska}
\email[]{Dorota.Gondek@obspm.fr}
\altaffiliation{also at Nicolaus Copernicus Astronomical Center, Bartycka 18,
00-716 Warszawa, Poland}
\author{Eric Gourgoulhon}
\email[]{Eric.Gourgoulhon@obspm.fr}
\affiliation{Laboratoire de l'Univers et de ses Th\'eories,
FRE 2462 du C.N.R.S., Observatoire de Paris, F-92195 Meudon Cedex, France}


\date{23 May 2002}

\begin{abstract}
We perform some numerical study of the secular triaxial instability
of rigidly rotating homogeneous fluid bodies in general relativity.
In the Newtonian limit, this instability arises at the bifurcation point
between the Maclaurin and Jacobi sequences. It can be driven in
astrophysical systems by viscous dissipation. We locate the onset of instability along several
constant baryon mass sequences of uniformly rotating axisymmetric
bodies for compaction parameter $M/R = 0-0.275$.
We find that general relativity weakens the Jacobi
like bar mode instability, but the stabilizing effect is not very
strong. According to our analysis the critical value of the ratio of
the kinetic energy to the absolute value of the gravitational
potential energy $(T/|W|)_{\rm crit}$ for compaction parameter as high
as $0.275$ is only $30\%$ higher than the Newtonian value. The
critical value of the eccentricity depends very weakly on the
degree of relativity and for $M/R=0.275$ is only $2\%$ larger
than the Newtonian value at the onset for the secular bar mode instability.
We compare our numerical results with recent analytical investigations
based on the post-Newtonian expansion.
\end{abstract}


\maketitle


\section{Introduction}

\subsection{The Maclaurin-Jacobi bifurcation point}

In Newtonian theory a self-gravitating incompressible fluid body
rotating at a moderate velocity around a fixed axis with respect to
some inertial frame takes the shape of a Maclaurin ellipsoid, which is
axisymmetric with respect to the rotation axis. For a higher rotation
rate, namely when the ratio of kinetic to gravitational potential
energy $T/|W|$ is larger than $0.1375$, another figure of equilibrium
exists: that of a Jacobi ellipsoid, which is triaxial
and rotates around its smallest axis \cite{Chand69}. Actually the Jacobi
ellipsoid is a preferred figure of equilibrium, since at fixed mass
and angular momentum, it has a lower total energy $E=T+W$
than a Maclaurin ellipsoid, due to its greater moment of
inertia $I$ with respect to the rotation axis.  Indeed, at fixed
angular momentum $J$, the kinetic energy
$T=J^2/(2 I)$ is a decreasing function of $I$, and for large values of
$J$, this decrease overcomes the effect of the gravitational
potential energy $W$, which increases with $I$.
Therefore, provided some mechanism acts for dissipating energy
while preserving angular momentum (for instance viscosity),
a Maclaurin ellipsoid with $T/|W|>0.1375$
will break its axial symmetry and migrate toward a Jacobi ellipsoid
\cite{PressT73,BertiR76,ChrisKST95}.
This is the secular ``bar mode'' instability
of rigidly rapidly rotating bodies. The qualifier {\em secular}
reflects the necessity of some dissipative mechanism to lower the
energy, the instability growth rate being controlled by the dissipation
time scale\footnote{For $T/|W|\ge 0.2738$, the Maclaurin spheroids are
subject to another instability, which is on the contrary {\em
dynamical}, i.e. it develops independently of any dissipative
mechanism and on a dynamical time scale (one rotation period).}.
As shown by Christodoulou et al.~\cite{ChrisKST95}, the
Jacobi-like bar mode instability appears only
if the fluid circulation is not conserved. If on the contrary,
the circulation is conserved (as in inviscid fluids submitted
only to potential forces), but not the angular momentum,
it is the Dedekind-like instability which develops instead.
The famous Chandrasekhar-Friedman-Schutz (CFS) instability
(see \cite{Anders02} for a review) belongs to this category.

The Jacobi-like bar mode instability, applied to neutron stars,
is particularly relevant to gravitational wave astrophysics.
Indeed a Jacobi ellipsoid has a time varying
mass quadrupole moment with respect to any inertial frame,
and therefore emits gravitational radiation, unlike a Maclaurin spheroid.
For a rapidly rotating neutron star, the typical frequency of gravitational
waves (twice the rotation frequency) falls in the bandwidth of
the interferometric detectors LIGO and VIRGO currently under
construction.

Neutron stars being highly relativistic objects,
the classical critical value $T/|W|=0.1375$, established for
incompressible Newtonian bodies, cannot a priori be applied to
them. The aim of the present article is thus to investigate
the effect of general relativity on the secular bar mode instability
of homogeneous incompressible bodies.
We do not discuss compressible fluids here. It has been
shown that compressibility has little effect on the triaxial
instability \cite{IpserM81}.

\subsection{Previous studies in the relativistic regime}

Chandrasekhar \cite{Chand67,Chand71} has examined the first order
post-Newtonian (PN) corrections to the Maclaurin and Jacobi
ellipsoids, by means of the tensor virial formalism.  This work has
been revisited recently by Taniguchi \cite{Tanig99}.
However, these authors have not computed the location of the
Maclaurin-Jacobi bifurcation point at the 1-PN level. This has
been done only recently by Shapiro \& Zane
\cite{ShapiZ98} and Di Girolamo \& Vietri \cite{GirolV02}.

On the numerical side, Bonazzola, Frieben \& Gourgoulhon
\cite{BonazFG96,BonazFG98} have investigated the secular
bar mode instability of rigidly rotating compressible stars
in general relativity. In the Newtonian
limit, they recover the classical result of James \cite{James64} (see
also \cite{SkinnL96}), namely that, for a polytropic equation of state,
the adiabatic index must be larger than $\gamma_{\rm crit} = 2.238$
for the bifurcation point to occur before the mass shedding limit
(Keplerian frequency). In the relativistic regime, they have shown
that general relativistic effects stabilize rotating stars against the
viscosity driven triaxial instability. In particular, they have found
that $\gamma_{\rm crit}$ is an increasing function of the stellar
compactness, reaching $\gamma_{\rm crit}\sim 2.8$ for a typical neutron
star compaction parameter.
This stabilizing tendency of general relativity has
been confirmed by the PN study of Shapiro \& Zane \cite{ShapiZ98} and
Di Girolamo \& Vietri \cite{GirolV02} mentioned above. Note that this behavior
contrasts with the CFS instability, which is strengthened by
general relativity \cite{StergF98,MorsiSB99}.

\subsection{The present work} \label{s:present}

In this paper, we improve the numerical technique over that
used by Bonazzola et al. \cite{BonazFG96,BonazFG98} by introducing
surface fitted coordinates, which enable us to treat the
density discontinuity at the surface of incompressible bodies.
Indeed the technique used in Refs.~\cite{BonazFG96,BonazFG98}
did not permit to compute any incompressible model. In particular,
it was not possible to compare the numerical results
in the Newtonian limit with the classical Maclaurin-Jacobi bifurcation
point. We shall perform such a comparison here. The very
good agreement obtained (relative discrepancy $\sim 10^{-6}$) provides
very strong support for the method we use for locating the
bifurcation point and which is essentially the same as that
presented in Ref.~\cite{BonazFG98}.

The plan of the paper is as follows. The analytical
formulation of the problem, including the approximations
we introduce, is presented in Sec.~\ref{s:basic}. Section~\ref{s:num}
then describes the numerical technique we employ, as well as
the various tests passed by the numerical code. The numerical
results are presented in Sec.~\ref{s:cal}, as well as a
detailed comparison with the PN studies \cite{ShapiZ98}
and \cite{GirolV02}. Finally, Sec.~\ref{s:summ} provides some summary
of our work.

\section{Basic assumptions and equations to be solved} \label{s:basic}

\subsection{Helical symmetry} \label{s:helic}

Let us consider a rotating star that is steadily increasing its
rotation rate, e.g. by accretion in a binary system.
Before the triaxial instability sets in, the spacetime generated by the rotating
star can be considered as {\em stationary} and {\em axisymmetric}, which
means that there exist two Killing vector fields, $\w{k}$ and $\w{m}$,
such that $\w{k}$ is timelike (at least far from the star) and $\w{m}$
is spacelike and its orbits are closed curves.

When the axisymmetry of the star is broken, the stationarity of spacetime
is also broken. In Newtonian theory, there is no
inertial frame in which a rotating triaxial object appears stationary,
i.e. does not depend upon the time. It can be stationary only in a
corotating frame, which is not inertial, so that the stationarity is
broken in this sense. The stationarity in the corotating frame
can be expressed geometrically by stating that the Newtonian spacetime
possesses a one-parameter symmetry group, whose integral curves
are helices. A generator of this symmetry group thus has the form
\begin{equation} \label{e:helical}
	\wg{\ell} = {\partial \over \partial t} + \Omega \,
	{\partial\over\partial \varphi} \ ,
\end{equation}
where $t$ and $\varphi$ are respectively the time and azimuthal
coordinates associated with an inertial observer, and $\Omega$
is the angular velocity with respect to the inertial frame.

In general relativity,
a rotating triaxial system cannot be stationary, even in the
corotating frame, as it radiates away
gravitational waves and therefore loses energy and angular momentum.
However, at the very point of the symmetry breaking, no gravitational
wave has yet been emitted. For sufficiently small deviations from
axisymmetry, we may neglect the gravitational radiation.
Therefore we shall assume that the spacetime has a helical symmetry,
as in the Newtonian case. The (suitably normalized) associated symmetry
generator $\wg{\ell}$ is then a Killing vector, which can be
written in the form (\ref{e:helical}) in weak-field regions (spacelike
infinity).
Note that spacetimes with helical symmetry have been also used
for describing binary systems with circular orbits
\cite{Detwe89,BonazGM97,BaumgCSST98,GourgGB02,FriedUS02}.

\subsection{Rigid rotation}

We model the stellar matter by a
perfect fluid, for which the stress-energy tensor takes the form
$ \w{T} = (e+p) \w{u} \otimes \w{u} + p \, \w{g} $,
where $\w{u}$ is the fluid 4-velocity, $e$ the fluid proper energy
density, $p$ the fluid pressure and $\w{g}$ the spacetime metric tensor.

A rigid motion is defined in relativity by the vanishing of the
expansion tensor
$\theta_{\alpha\beta} := (g_\alpha^{\ \mu}+u_\alpha u^\mu)
	(g_\beta^{\ \nu}+u_\beta u^\nu)
	\nabla_{(\nu} u_{\mu)}$ of the 4-velocity $\w{u}$. In
presence of the Killing vector $\wg{\ell}$, this can be realized by requiring
the colinearity of $\w{u}$ and $\wg{\ell}$ (supposing that
the fluid occupies only the region where $\wg{\ell}$ is timelike) :
\begin{equation} \label{e:rigid}
	\w{u} = \lambda \, \wg{\ell} \ ,
\end{equation}
where $\lambda$ is a scalar field related to the norm of $\wg{\ell}$ by
the normalization of the 4-velocity
$\lambda = (-\wg{\ell}\cdot\wg{\ell})^{-1/2}$.

For a perfect fluid at zero temperature, the momentum-energy
conservation equation $\nabla\cdot \w{T}=0$ can be recast as
\cite{Lichn67,Carte79}
\begin{equation}
\label{e:canon}
\w{u} \cdot (\nabla \wedge \wg{\pi}) = 0
\end{equation}
\begin{equation}
\label{e:conserv}
\nabla \cdot (n \w{u}) = 0 \ ,
\end{equation}
where $n$ is the proper baryon number density and $\wg{\pi}$ is
the momentum 1-form  $\wg{\pi} := h \w{u}$,
$h$ being the fluid specific enthalpy: $h := (e+p) / (m_{\rm B} n)$,
where $m_{\rm B}$ is some mean baryon mass.
In Eq.~(\ref{e:canon}), $\nabla \wedge \wg{\pi}$  denotes the exterior
derivative of $\wg{\pi}$, the so-called {\em vorticity 2-form} \cite{Lichn67}.
For the rigid motion we are considering, the baryon number conservation
equation (\ref{e:conserv}) is automatically satisfied, thanks to
the colinearity of $\w{u}$ with the symmetry generator $\wg{\ell}$.
Moreover, the equation of motion~(\ref{e:canon}) can be reduced to
a first integral. Indeed Cartan's identity applied to the
Lie derivative of the 1-form $\wg{\pi}$ along the vector field $\wg{\ell}$
leads to
\begin{equation} \label{e:Cartan}
	\pounds_{\wg{\ell}} \wg{\pi} = \wg{\ell}\cdot(\nabla\wedge\wg{\pi})
	+ \nabla(\wg{\ell}\cdot\wg{\pi}) = 0 \ ,
\end{equation}
where the second equality holds thanks to the helical symmetry:
 $\pounds_{\wg{\ell}} \wg{\pi} = 0$.
Inserting relation (\ref{e:rigid}) into the equation of fluid motion
(\ref{e:canon})
shows that the first term in Eq.~(\ref{e:Cartan}) vanishes identically,
so that one gets the first integral of motion \cite{Carte79}
\begin{equation} \label{e:int_prem_rigid}
	\wg{\ell} \cdot \wg{\pi} = - \lambda^{-1} h = {\rm const.}
\end{equation}
In the axisymmetric and stationary case, where $\wg{\ell}$
is a linear combination of the two Killing vectors [Eq.~(\ref{e:ell_axi})
below], one recovers the classical expression \cite{Boyer65}.

The first integral (\ref{e:int_prem_rigid}) can be re-expressed in terms of
the 3+1 formalism of general relativity. Let us introduce the spacetime foliation
by the $t={\rm const}$ hypersurfaces $\Sigma_t$ and the associated
future-directed unit vector field $\w{n}$ everywhere normal to $\Sigma_t$.
$\w{n}$ is the 4-velocity of the so-called {\em Eulerian observer} (also
called {\em ZAMO} or {\em locally non-rotating observer}).
We have the following orthogonal split of the fluid 4-velocity:
\begin{equation}  \label{e:u_ortho}
	\w{u} = \Gamma (\w{n} + \w{U}) \qquad \mbox{with} \quad
		\w{n}\cdot\w{U} = 0 \ ,
\end{equation}
with the Lorentz factor
\begin{equation} \label{e:gamma}
	\Gamma = - \w{n} \cdot \w{u} = (1-\w{U}\cdot\w{U})^{-1/2} \ .
\end{equation}
The spacelike vector $\w{U}$ is the fluid 3-velocity as measured by
the Eulerian observer and the second equality in the above equation
results from the normalization of the 4-velocity $\w{u}$.
Similarly, we have an orthogonal split of
the helical Killing vector:
\begin{equation}  \label{e:ell_ortho}
	\wg{\ell} = N \w{n} + \w{B} \qquad \mbox{with} \quad
		\w{n}\cdot\w{B} = 0 \ ,
\end{equation}
where $N$ is the lapse function, governing the proper time evolution
between two neighboring hypersurfaces $\Sigma_t$.
From relation (\ref{e:rigid}) and
the two orthogonal decompositions (\ref{e:u_ortho}) and
(\ref{e:ell_ortho}), we get
$\lambda = \Gamma/N$, so that the (logarithm of the) first integral
(\ref{e:int_prem_rigid}) can be written
\begin{equation} \label{e:int_prem}
	H + \nu - \ln \Gamma = {\rm const.}
\end{equation}
with
\begin{equation} \label{e:Hnu_def}
	H := \ln h \qquad \mbox{and} \qquad \nu := \ln N \ .
\end{equation}
In the non-relativistic limit, $H$ tends toward the classical
specific enthalpy (excluding the rest-mass energy) of the fluid,
whereas $\nu$ tends toward the Newtonian gravitational potential.
Therefore, in the Newtonian limit, where
$\ln\Gamma \rightarrow \w{U}^2/2$ and
$\w{U} \rightarrow \wg{\Omega}\times \w{r}$, Eq.~(\ref{e:int_prem}) reduces to
the classical first integral of motion
\begin{equation}
	H + \nu - {1\over 2} (\wg{\Omega}\times \w{r})^2 = {\rm const.}
\end{equation}

\subsection{Einstein equations}

When the rigidly rotating star is still stationary and axisymmetric,
it is well known that a coordinate system
$x^\mu=(t,r,\theta,\varphi)$ can
be chosen so that the metric takes the Papapetrou form (see e.g.
\cite{BonazGSM93} and references therein)
\begin{equation}  \label{e:metric_axi}
g_{\mu\nu}\, dx^\mu dx^\nu  =
-N^2 \, {\rm d}t^2 + B^2r^2\sin^2\theta
({\rm d}\varphi - N^\varphi {\rm d}t)^2
 + A^2 ({\rm d}r^2 + r^2 {\rm d}\theta^2)  \ ,
\end{equation}
where $N$, $N^\varphi$, $A$ and $B$ are four functions of $(r,\theta)$.
The coordinate vectors $\partial/\partial t$ and $\partial/\partial\varphi$
are then the two Killing vectors $\w{k}$ and $\w{m}$ mentioned in
Sec.~\ref{s:helic}. The helical Killing vector $\wg{\ell}$ is expressible as
\begin{equation} \label{e:ell_axi}
	\wg{\ell} = \w{k} + \Omega \w{m} \ .
\end{equation}
With the form (\ref{e:metric_axi}), the Einstein equations reduce
to a set of four coupled elliptic equations (see e.g.
\cite{GourgHLPBM99} for the precise form).

When the triaxial instability sets in, we shall consider that the
metric is a perturbation of (\ref{e:metric_axi}), which we write
as \cite{BonazFG98}
\begin{equation}  \label{e:metric_3d}
g_{\mu\nu}\, dx^\mu dx^\nu  =
-N^2 \, {\rm d}t^2 + B^2 r^2\sin^2\theta
({\rm d}\varphi - N^\varphi {\rm d}t)^2
 + A^2 ({\rm d}r - N^r {\rm d}t)^2
 + r^2 A^2 ({\rm d}\theta - N^\theta {\rm d}t)^2  \ ,
\end{equation}
where $N$, $N^r$, $N^\theta$, $N^\varphi$, $A$ and $B$ are six functions
of $(r,\theta,\varphi')$, with
\begin{equation}
	\varphi' := \varphi - \Omega t \ .
\end{equation}
Note that $\partial/\partial t$
and $\partial/\partial\varphi$ are no longer Killing vectors, only
$\wg{\ell}$ remains Killing.
The coordinate system $x^{\mu'}=(t,r,\theta,\varphi')$ is
adapted to the Killing vector $\wg{\ell}$
and $t$ is an ignorable coordinate in this system. We
call $x^{\mu'}$ the {\em corotating} coordinate system, and
$x^\mu$ the {\em nonrotating} one.
$N$ is the lapse function already introduced in
Eq.~(\ref{e:ell_ortho}).
$N^i=(N^r,N^\theta,N^\varphi)$ is (minus\footnote{In numerical relativity,
the current convention is to call {\em shift vector} the 3-vector
$\beta^i = - N^i$.}) the {\em shift vector} of the nonrotating
coordinate system.

In the triaxial case, there is no equivalent of the Papapetrou theorem
\cite{Papap66,KundtT66,Carte69} which, in absence of convective
motions (in the meridional planes), allows one to set to zero all the
off-diagonal components of the metric tensor of
axisymmetric and stationary spacetimes, except for $g_{t\varphi}$.
So, in principle, all the metric components should be non-vanishing
in Eq.~(\ref{e:metric_3d}). However, we retained only $g_{tr}$ and
$g_{t\theta}$ as the extra non zero components with respect to the
axisymmetric case. We did so as an approximation in order to simplify
the writing of Einstein equations. This approximation can be justified
by the following remarks: (i) the metric element (\ref{e:metric_3d}) is
exact in the stationary axisymmetric case, (ii) it encompasses the
first order PN metric. This means that the
neglected terms are of second order PN and
moreover vanish for stationary axisymmetric rotating stars. We consider these terms
to be negligible in our study of the verge of the non-axisymmetric
instability.

The Einstein equation leads to the following set of partial differential
equations:
\begin{eqnarray}
& & \underline \Delta \, \nu \; = \;
   4\pi A^2 ( E + 3 p + (E+p) U_i U^i )
   + A^2 K_{ij} K^{ij}
   - \overline\nabla_i \nu \, \overline\nabla^i (\nu + \beta)
   					\label{e:Einstein1} \\
& & \underline\Delta \, N^i  +\frac{1}{3} \overline\nabla^i
	\overline\nabla_j N^j  \; = \;
   - 16\pi NA^2 (E+p) U^i
	 + N B^{-2} K^{ij} \overline\nabla_j(6\beta-\nu)  \label{e:Einstein2}	\\
& & \Delta_2 \, [ (NB-1) \, r\sin\theta]  \; = \;
    16 \pi NA^2 B p \, r \sin\theta    	\label{e:Einstein3} \\
& & \Delta_2 \, \zeta   \; = \;
  8\pi  A^2 \, [p + (E+p)U_i U^i]
   + \frac{3}{2} A^2 K_{ij} K^{ij}
	 - \overline\nabla_i \nu \overline\nabla^i \nu   \ , \label{e:Einstein4}
\end{eqnarray}
where the following notations have been introduced
[see also Eq.~(\ref{e:Hnu_def})]
\begin{equation}
  \zeta :=\ln (AN) \ , \quad \beta := \ln B \ ,
\end{equation}
and $\overline\nabla_i$ denotes the covariant derivative with
respect to the flat 3-metric $f_{ij} = {\rm diag}(1,r^2,r^2\sin^2\theta)$,
$\underline\Delta = \overline\nabla_i \overline\nabla^i$ the corresponding
Laplacian and $\Delta_2$ is the Laplacian in the 2-dimensional
space spanned by $(r,\theta)$:
\begin{equation}
   \Delta_2  := \frac{\partial^2}{\partial r^2}
   	+ \frac{1}{r}  \frac{\partial}{\partial r}
   	+ \frac{1}{r^2}  \frac{\partial^2}{\partial \theta^2} \ .
\end{equation}
In Eqs.~(\ref{e:Einstein1})-(\ref{e:Einstein4}), $U^i$ denotes the
spatial components of the fluid 3-velocity $\w{U}$
[cf. Eq.~(\ref{e:u_ortho})]. They can be expressed as
\begin{equation} \label{e:U_i}
	U^r = - \frac{N^r}{N} , \quad
	U^\theta = - \frac{N^\theta}{N} , \quad
	U^\varphi = \frac{1}{N}(\Omega - N^\varphi) , \quad	
\end{equation}
and $\Gamma$ is deduced from $\w{U}$ by means of Eq.~(\ref{e:gamma}):
\begin{equation}
	\Gamma = \left( 1 - A^2 [(U^r)^2 + r^2 (U^\theta)^2]
		- B^2 r^2 \sin^2\theta (U^\varphi)^2 \right) ^{-1/2} \ .
\end{equation}
We have also introduced the energy density $E$ as measured by the
Eulerian observer:
\begin{equation} \label{e:ener_euler}
	E = \w{n}\cdot\w{T}\cdot\w{n} = \Gamma^2 (e+p) - p \ .
\end{equation}
Another quantity not yet defined and which appears in
Eqs.~(\ref{e:Einstein1})-(\ref{e:Einstein4}) is the extrinsic
curvature tensor $\w{K}$ of the hypersurface $\Sigma_t$.

\subsection{Equation of state}

The above equations  must be supplemented by
an equation of state (EOS), i.e. a relation between $H$,
which appears in the first integral (\ref{e:int_prem}), and $e$ and $p$
which appear in the source terms of the Einstein
equations (\ref{e:Einstein1})-(\ref{e:Einstein4}), via
Eq.~(\ref{e:ener_euler}).
For the incompressible matter we are considering, we choose this EOS
to be
\begin{eqnarray}
	e(H) &=& \rho_0  \label{e:eos_e} \\
	p(H) &=& \rho_0 (\exp(H) - 1 ) \label{e:eos_p} \\
	n(H) &=& \rho_0 / m_{\rm B} \ ,
\end{eqnarray}
where $\rho_0$ is a constant, representing the constant proper energy
density of the fluid.

\section{Numerical code and its tests} \label{s:num}

\subsection{Numerical technique} \label{s:num_tech}

The basic idea is to solve the equations presented in
Sec.~\ref{s:basic} by means of an iterative scheme to
get an axisymmetric stationary solution, and then to
introduce some triaxial perturbation and resume the iterative
scheme. If the perturbation is damped (resp. grows)
as long as the iteration proceeds, the equilibrium
configuration will be declared stable (resp. unstable).
This procedure has been used already in
the works~\cite{BonazFG96,BonazFG98}.
We shall prove here, by comparison with the analytical
result, that it correctly locates the secular
instability point along the Maclaurin sequence.

We solve the system of non-linear elliptic equations
(\ref{e:Einstein1})-(\ref{e:Einstein4}) by means of the multi-domain
spectral method presented in Ref.~\cite{BonazGM98}.
The nice feature of this method is that it introduces
surface-fitted coordinates $(\xi,\theta',\varphi')$,
so that the density discontinuity at the stellar surface is
exactly located at the boundary between two domains.
In this way, all the fields are $C^\infty$ functions
in each domain. This avoids any spurious oscillations
(Gibbs phenomenon) and
results in a very high precision.
As discussed in Sec.~\ref{s:present}, this technique
constitutes the major improvement with respect to the
numerical method used in Ref.~\cite{BonazFG98}.
Another difference with Ref.~\cite{BonazFG98} is that
we employ the technique presented in Ref.~\cite{GrandBGM01}
to solve the vector elliptic equation (\ref{e:Einstein2})
for the shift vector. In particular, we solve for the
Cartesian components of the shift vector, whereas
\cite{BonazFG98} solved for the spherical components.

A weak point of the surface-fitted coordinate technique of
Ref.~\cite{BonazGM98} is that the transformation from
the computational coordinates $(\xi,\theta',\varphi')$
to the physical ones $(r,\theta,\varphi)$ becomes
singular when the ratio of polar to equatorial
radius $r_{\rm p}/r_{\rm eq}$ is lower than $\sim 1/3$.
This prevents us from computing very flattened configurations,
but fortunately this causes no trouble in
reaching the Maclaurin-Jacobi
bifurcation point, which is at $r_{\rm p}/r_{\rm eq}\simeq 0.58$.

\begin{figure}
\includegraphics[height=6cm]{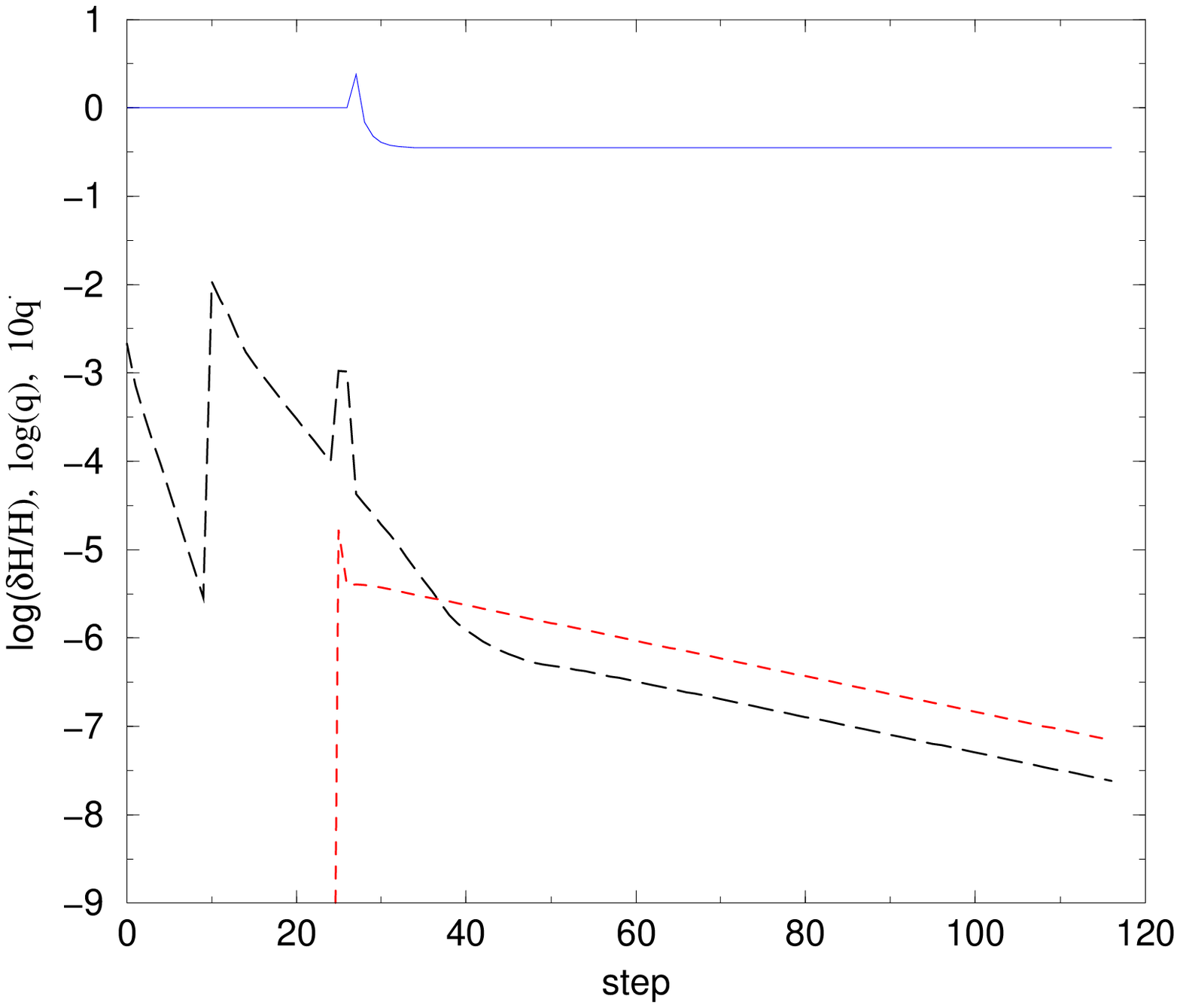} \qquad
\includegraphics[height=6cm]{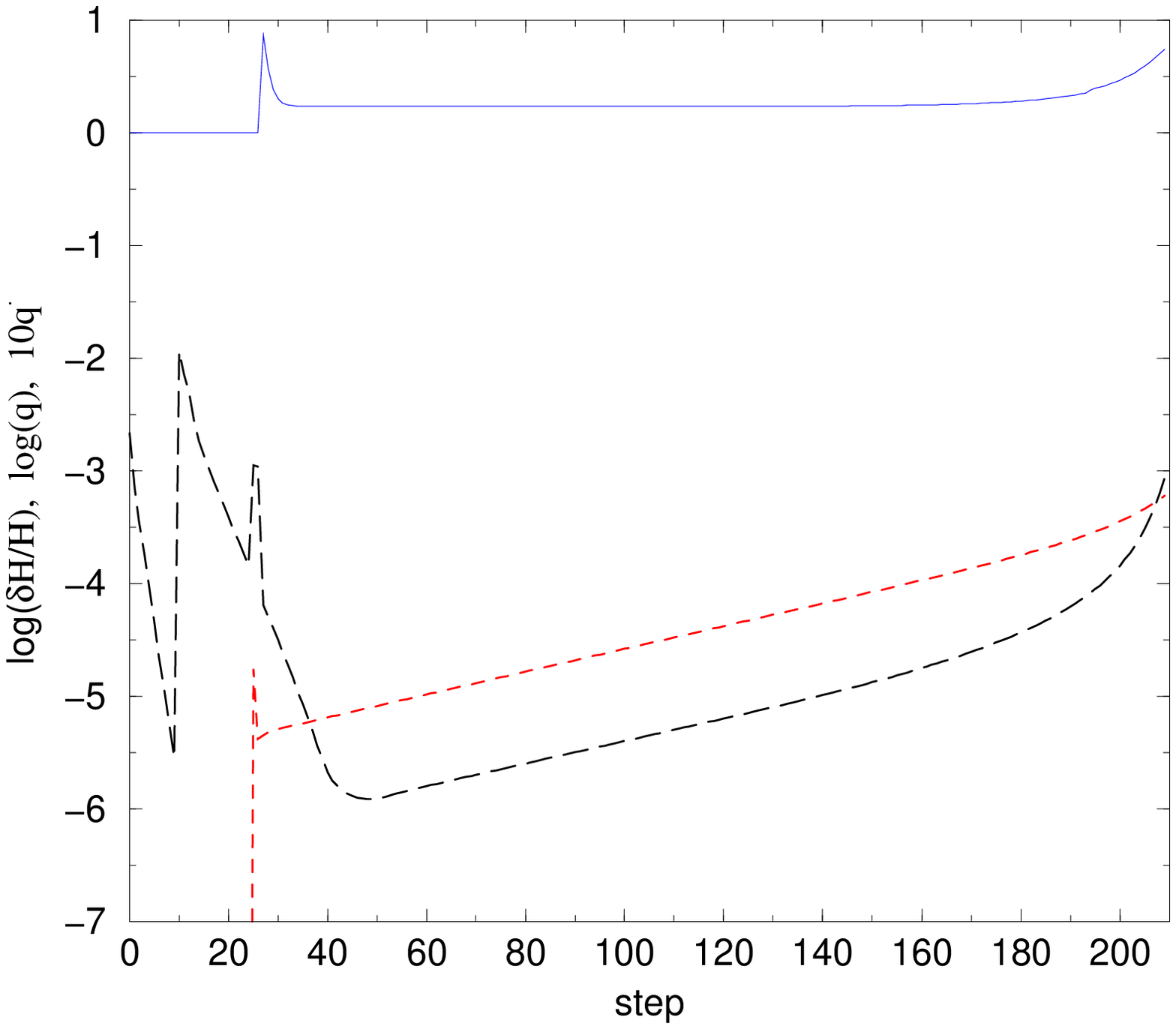}
\caption{\label{f:evol_q} Evolution during the iterative procedure of
the convergence indicator $\delta H/H$ (long-dashed line), the
triaxial perturbation $q$ (short-dashed line) and its growth rate
$\dot q$ (solid line). The left figure corresponds to a stable
configuration with respect to nonaxisymmetric perturbation, the right one an
unstable one. $\delta H/H$ and $q$ are depicted in logarithmic units,
while $\dot q$ is multiplied by 10 for better readability.
The discontinuity at step 10 is due to the switch of the
rotation and that at step 25 to the switch of the triaxial
perturbation.}
\end{figure}

The iterative procedure is as follows.
First one must set a
value of the central enthalpy $H_c$ and the rotation
frequency $\Omega$, in order to pick out a unique
rotating axisymmetric configuration.
The iteration is then started from very crude values:
flat spacetime and spherical stellar shape.
We solve the Einstein equations (\ref{e:Einstein1})-(\ref{e:Einstein4})
with the spherical energy density and pressure distribution
in their right-hand side. We then plug the obtained value of
$\nu$ in the first integral of motion (\ref{e:int_prem}), along
with $H_c$, to get the enthalpy field $H$. The zero of this field
defines the new stellar surface to which we adapt the
computational coordinates $(\xi,\theta',\varphi')$
(see \cite{BonazGM98} for details).
From $H$ we compute new values of the energy density and
pressure according to the EOS (\ref{e:eos_e})-(\ref{e:eos_p}).
These values are then put on the right-hand side of the
Einstein equations (\ref{e:Einstein1})-(\ref{e:Einstein4})
and a new iteration begins.
Usually the rotation velocity is set to zero for the ten
first steps. It is then switched on, either integrally
or (for very rapidly rotating configurations) gradually.
The convergence of the procedure is monitored by
computing the relative difference $\delta H/H$ between
the enthalpy fields at two successive steps
(long-dashed line in Fig.~\ref{f:evol_q}). The iterative
procedure is stopped when $\delta H/H$ goes below a certain
threshold, typically $10^{-7}$, or for high precision
computations $10^{-12}$.

\begin{figure}
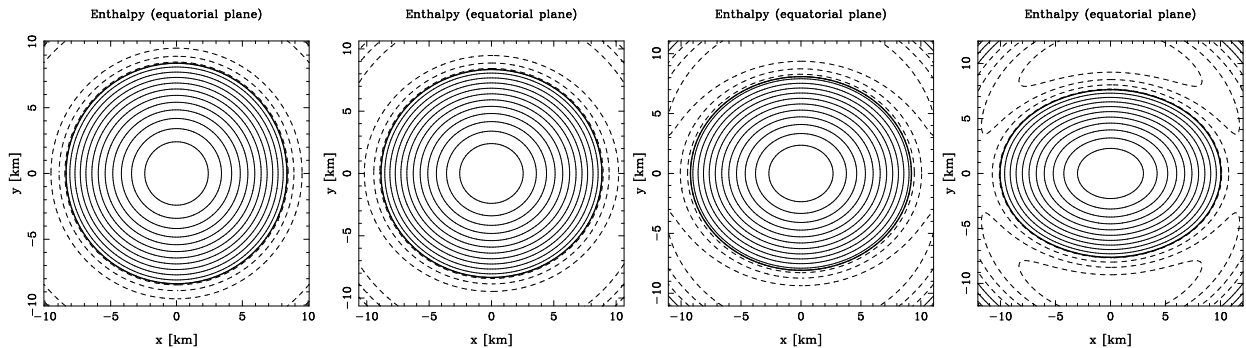

\includegraphics[height=4.5cm]{evolxy1.eps}
\includegraphics[height=4.5cm]{evolxyn150.eps}
\includegraphics[height=4.5cm]{evolxyn180.eps}
\includegraphics[height=4.5cm]{evolxyn200.eps}
\caption{\label{f:bar} Isocontours of the enthalpy $H$
in the equatorial plane at steps 20, 150,
 180 and 200 for the unstable model presented in
Fig.~\ref{f:evol_q}. Dashed lines denote negative values of $H$,
which correspond to the exterior of the star. The thick solid line
denotes the stellar surface.}
\end{figure}

After this convergence has been achieved,
we switch on a triaxial $m=2$ perturbation
by modifying the metric potential $\nu$ according to
\begin{equation}
	\nu \rightarrow \nu  \left( 1 + \varepsilon \sin^2\theta
		\cos 2\varphi \right) \ ,
\end{equation}
where $\varepsilon$ is a small constant, typically
$\varepsilon \simeq 10^{-5}$ to $10^{-3}$.
Then we continue the iteration procedure as described above,
without any further modification of the equations.
At each step, we evaluate the quantity
\begin{equation} \label{e:max_triax}
	q := \max \sqrt{{\hat\nu_2}^2 + {\hat\nu_3}^2 } \ ,	
\end{equation}
where $\hat\nu_2$ and $\hat\nu_3$ are the $m=2$
coefficients of the Fourier expansion of the $\varphi$
part of $\nu$:
\begin{equation}
	\nu(\varphi) = \hat\nu_0 + \hat\nu_2\cos(2\varphi)
		+ \hat\nu_3\sin(2\varphi)
		+ \hat\nu_4\cos(4\varphi)
		+ \hat\nu_5\sin(4\varphi)
		+ \cdots
\end{equation}
and the $\max$ in Eq.~(\ref{e:max_triax}) is taken over all
the $r$ and $\theta$ coefficients.
$q$ is used to monitor the evolution of the triaxial
perturbation: if $q\rightarrow 0$ as the iteration proceeds,
we conclude that the perturbation decays and the axisymmetric
configuration is a stable one.
In the vicinity of the marginally stable configuration, the
decay or growth of the perturbation turns out to be pretty
small. In order to facilitate the diagnostic, we monitor
instead the {\em relative} growth rate of the $m=2$ part,
defined as
\begin{equation}
	\dot q := \frac{q - q_{\rm previous\ step}}{q_{\rm previous\  step}} \ .
\end{equation}
After some transitory regime, $\dot q$ turned out to be constant.
If  $\dot q > 0$ (resp. $\dot q < 0$) we conclude that the configuration is
unstable (resp. stable).
The behaviors of $\delta H/H$, $q$ and
$\dot q$ in two typical computations are shown in Fig.~\ref{f:evol_q},
whereas Fig.~\ref{f:bar} depicts the development of the triaxial
instability.

Note that the above method does not require to specify
some value of viscosity. Whatever this value,
the effect of viscosity is simulated by the rigid rotation
profile that we impose at each step in the iterative
procedure via Eq.~(\ref{e:U_i}).
If we consider the iteration as mimicking some time evolution,
this means that the time elapsed between two successive steps
has been long enough for the actual viscosity to have rigidified
the fluid flow. Consequently, a quantity that we cannot
get by our method is the instability time scale.
As recalled in the Introduction, this
time scale depends upon the actual value of viscosity, but not
the instability itself.

The numerical code implementing the above procedure has
been constructed upon the C++ library {\sc Lorene} \cite{lorene}.
Numerical computations have
been performed on SGI Origin200 as well as Linux PC workstations.
Three domains have been used, the innermost one corresponding to
the interior of the star, and the outermost one extending to
spacelike infinity by means of a suitable compactification
(see \cite{BonazGM98} for details).
The number of spectral coefficients used in each
domain is $N_r\times N_\theta \times N_\varphi = 33 \times 17 \times 4$.
The corresponding memory requirement is 40 MB and a typical CPU time
(e.g. corresponding to the first 60 steps of Fig.~\ref{f:evol_q},
which are sufficient to conclude about the stability of the configuration)
is 3 min on an Intel Pentium IV 1.5~GHz processor.

\subsection{Test of the numerical code}

Numerous tests have been performed to assess the validity
of the method and the accuracy of the numerical code.
We present here successively tests for axisymmetric configurations
in general relativity
and tests about the determination of the triaxial instability point
in the Newtonian limit. Tests regarding the triaxial
instability in the relativistic case are deferred to
Sec.~\ref{s:resu_inst},
where we present a detailed comparison with analytical
(post-Newtonian) results.

\subsubsection{Tests in the axisymmetric regime}

The multi-domain spectral technique with surface-fitted coordinates
has already
been tested in the Newtonian regime, giving a rapidly
rotating Maclaurin ellipsoid with a relative error of the order of
$10^{-12}$ \cite{BonazGM98}. It has been also shown in
Ref.~\cite{BonazGM98} that the error is evanescent, i.e. that
it decays exponentially with the number of spectral coefficients,
which is typical of spectral methods.

The accuracy of the computed relativistic axisymmetric models is
estimated using two general relativistic virial identities GRV2
\cite{Bonaz73,BonazG94} and GRV3 \cite{GourgB94}. These two virial
error indicators are integral identities which must be satisfied by
any solution of the Einstein equations and which are not imposed
during the numerical procedure (see Ref.~\cite{NozawSGE98} for the
computation of GRV2 and GRV3). GRV3 is a generalization to general
relativity of the classical virial theorem.
We have obtained values of GRV2 and GRV3 of the
order of $10^{-12}$ in Newtonian regime and $10^{-8}-10^{-4}$ for high
rotation and high compaction parameter. The virial errors are at least
one order of magnitude better than obtained in the previous code used
for calculating rapidly rotating strange stars \cite{GourgHLPBLM99,%
GondeBZGRDD00, GondeHZG00, ZduniBKHG00, ZduniHGG00, %
GondeSBKG01, GondeBKZG01, AmsteBGK02}.

\begin{table}
\caption{\label{t:ansorg} Comparison with the numerical results of
Ansorg et al. \cite{AnsorKM02} for a highly relativistic stationary
axisymmetric model $p_{\rm c}/\rho_0=1$ ($H_c=\ln 2$) and
$r_{\rm p}/r_{\rm eq} =0.7$,
with compaction parameter ${M}/R_{\rm circ} \sim 0.39$. Meaning
of the symbols [in geometrized units ($G=c=1$)] are as follows:
dimensionless angular velocity $\overline\Omega := \Omega/\rho_0^{1/2}$,
gravitational mass $\overline M:=M\rho_0^{1/2}$, baryon mass
$\overline M_{\rm B}:=M_{\rm B}\rho_0^{1/2}$, circumferential equatorial radius
$\overline R_{\rm circ} := R_{\rm circ}\rho_0^{1/2}$, angular momentum
$\overline J:=J\rho_0$. $Z_{\rm p}$, $Z^{\rm f}_{\rm eq}$, and
$Z^{\rm b}_{\rm eq}$ are the redshifts in the polar direction,
in the equatorial plane along the direction of motion and opposite to
that direction respectively.}
\begin{ruledtabular}
\begin{tabular}{lll}
   & our value & relative diff. \\
\hline
$\overline\Omega $ & 1.4116964 & 0.0009 $\%$ \\
$\overline M$ & 0.1358182 & 0.015 $\%$  \\
$\overline M_0$     & 0.1863648 & 0.014 $\%$  \\
$\overline R_{\rm circ}$ & 0.34548954 & 0.0039 $\%$ \\
$\overline J$       & 0.01405836 & 0.0017 $\%$ \\
$Z_{\rm p}$     & 1.7073708 &  0.001 $\%$ \\
$Z^{\rm f}_{\rm eq}$ & $-0.1625613$ & 0.017 $\%$ \\
$Z^{\rm b}_{\rm eq}$ & 11.3539974  & 0.00073 $\%$ \\
\end{tabular}
\end{ruledtabular}
\end{table}

Another test in the strongly relativistic regime is the comparison
with the highly accurate recent code of Ansorg et
al. \cite{AnsorKM02}, also based on spectral methods but using
very different coordinates and resolution scheme.
The relative differences defined as ${\rm
diff}={\rm |(AKM - GG)/AKM|} $ (where AKM and GG are the values
obtained by Ansorg et al. \cite{AnsorKM02} -- their Table~1 --
and us respectively),
are presented in
Table~\ref{t:ansorg} for an incompressible rapidly rotating body with
a compaction parameter $M/R_{\rm circ} =0.393$, where
$R_{\rm circ}$ is the circumferential radius of the star.
Note that this model is much more relativistic than any realistic
neutron star, so that we are really testing the strong field
behavior of our code. Table~\ref{t:ansorg} shows that the
agreement with Ansorg et al. is very good, of the order of
$10^{-4}$ or better. For this model,
the GRV2 and GRV3 errors from our code
are $2\, 10^{-5}$ and $3\, 10^{-6}$ respectively.

\subsubsection{Tests about the code capability to find the
triaxial instability}
Widely used indicators for the onset of
instability are the eccentricity $e$ and
the ratio of the kinetic energy to the absolute value of
the gravitational potential energy $T/|W|$, defined
respectively as
\begin{eqnarray}
e^2 &=&1-(r_{\rm p}/r_{\rm eq})^2 \label{eq:ecc} \ , \\
T/|W|&=&{{\Omega J/2}\over{\Omega J/2 + M_{\rm B} -M}} \ , \label{e:tsw}
\end{eqnarray}
where $r_{\rm p}$ and $r_{\rm eq}$ are the polar and equatorial
coordinate radius, $J$ the total angular momentum and $M_{\rm B}$ the
baryon mass.  Note that $T/|W|$ is gauge invariant for uniformly
rotating incompressible objects\footnote{For compressible bodies,
$T/|W|$ can be defined according to Eqs.~(20)-(22) of
Ref.~\cite{FriedIP86}, for which $M_{\rm B}$ at the denominator of
Eq.~(\ref{e:tsw}) must be replaced by some ``proper'' internal energy
$M_{\rm p}$; in general relativity this last quantity has less
physical meaning than the total baryon mass $M_{\rm B}$.}, while
eccentricity is coordinate dependent.  In the Newtonian regime, we
have employ the code to locate the bifurcation point between
the Maclaurin sequence
and the Jacobi one, and found it to be at 
\begin{equation}
({T}/{|W|})_{\rm crit} = 0.137526 \quad \mbox{and}
\quad e_{\rm crit}=0.812667 \ .
\end{equation}
Both quantities differ from
the exact value $(T/|W|)_{\rm crit,Newt}$ and $e_{\rm crit,Newt}$
\cite{Chand69} by only $10^{-6}$.

\begin{figure}
\includegraphics[height=10cm]{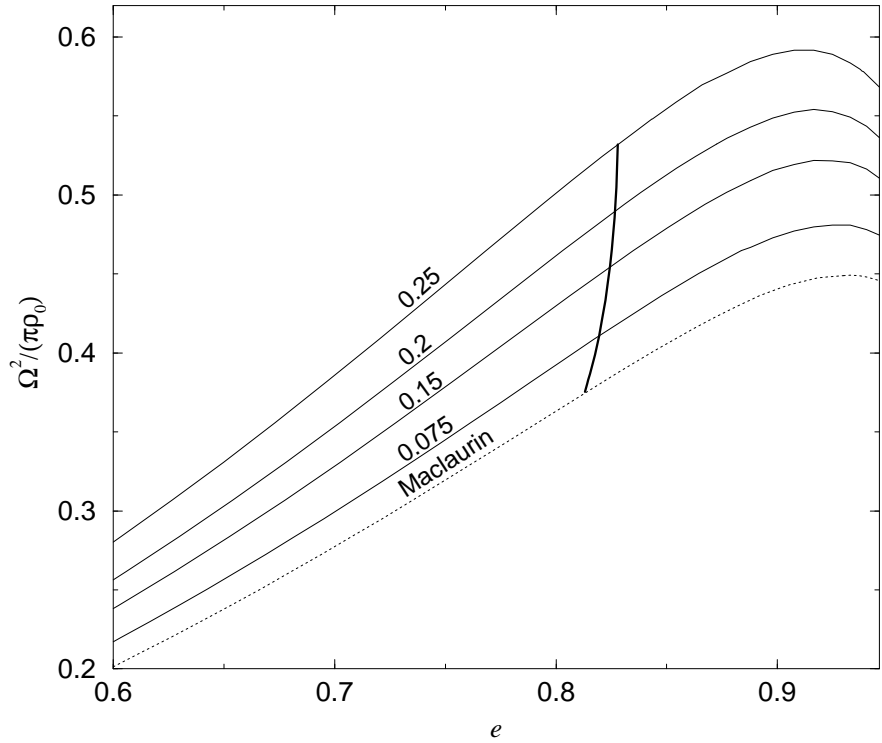}
\caption{\label{f:omeg_ecc_ours}
Square of the angular velocity
(in units of  $\pi G \rho_0$) as a function of the eccentricity [defined
by Eq.~(\ref{eq:ecc})] for axisymmetric equilibrium sequences of
constant rest mass. Each sequence is labeled by the compaction
parameter $M_{\rm s}/R_{\rm s}$ of its nonrotating member.
The dotted line denotes the Maclaurin sequence
($M_{\rm s}/R_{\rm s}=0$).  The thick line connects the secular instability
points.}
\end{figure}

\begin{figure}
\includegraphics[height=10cm]{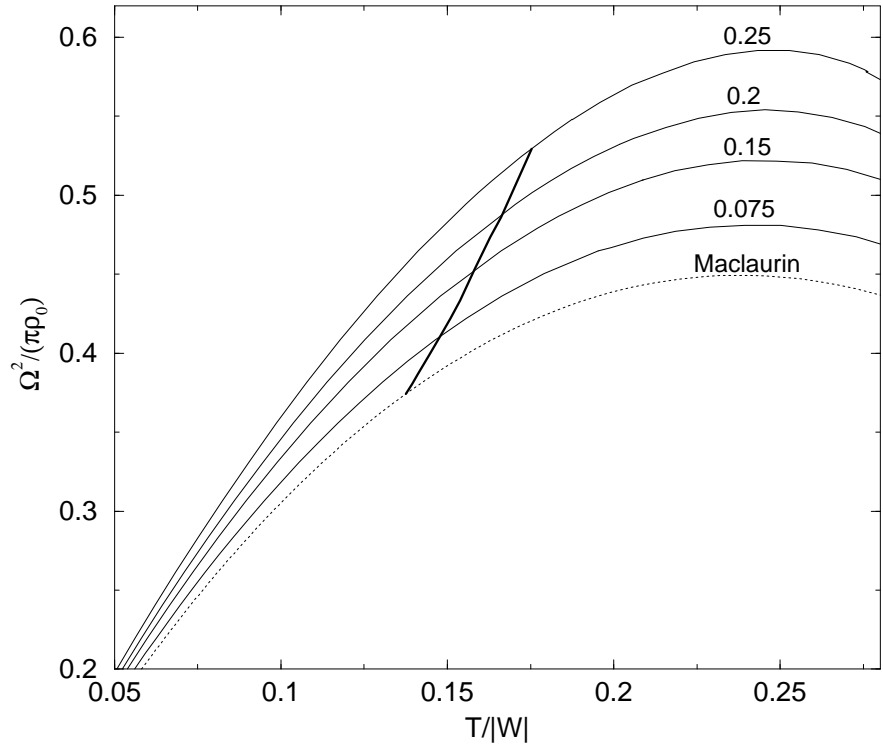}
\caption{\label{f:omeg_tsw_ours} Square of the angular velocity
versus the ratio of the kinetic energy to the absolute value of the
gravitational potential energy [defined by Eq.~(\ref{e:tsw})] for
the same sequences as in Fig.~\ref{f:omeg_ecc_ours}. The thick line
connects the secular instability points.}
\end{figure}

\begin{figure}
\includegraphics[height=10cm]{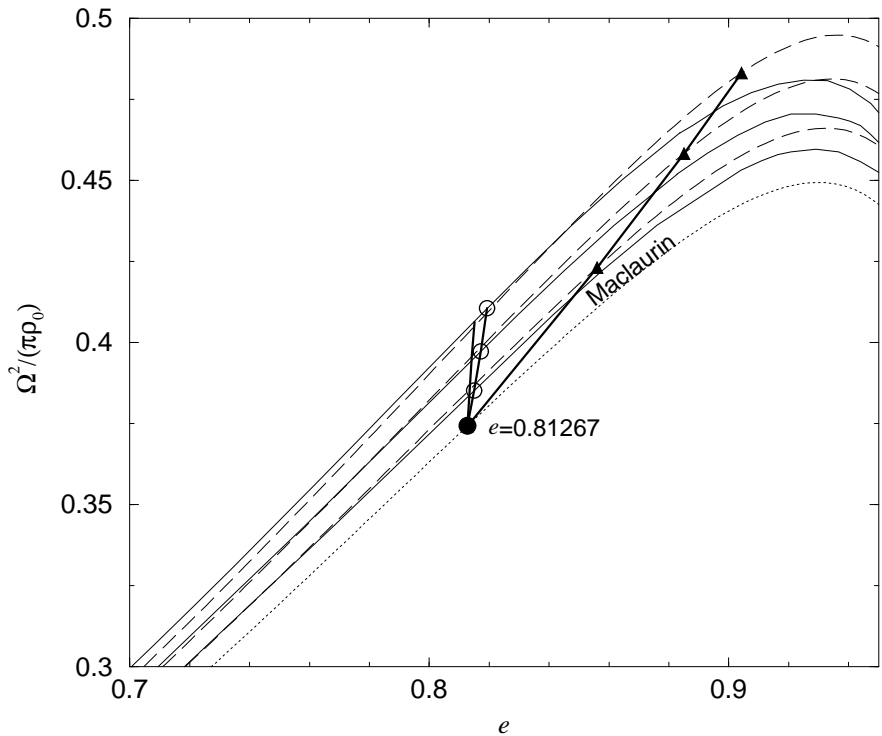}
\caption{\label{f:omeg_ecc} Comparison between the present relativistic
computations of the bar mode instability points (thick solid
line with circles) and that of two different PN calculations
performed by \cite{ShapiZ98} (thick solid line with triangles) and by
\cite{GirolV02} (thick solid line). The Newtonian Maclaurin-Jacobi
bifurcation point is shown as a filled circle.  Thin solid lines and
long-dashed lines corresponds respectively to fully relativistic
calculations and to the ellipsoidal PN ones of \cite{ShapiZ98}
for axisymmetric equilibrium sequences with compaction parameter $M_{\rm s}/R_{\rm s}$
= 0.075; 0.05; 0.025 from top to bottom. The dotted line represents the
Newtonian Maclaurin sequence ($M_{\rm s}/R_{\rm s}=0$).
}

\end{figure}

\begin{figure}
\includegraphics[height=10cm]{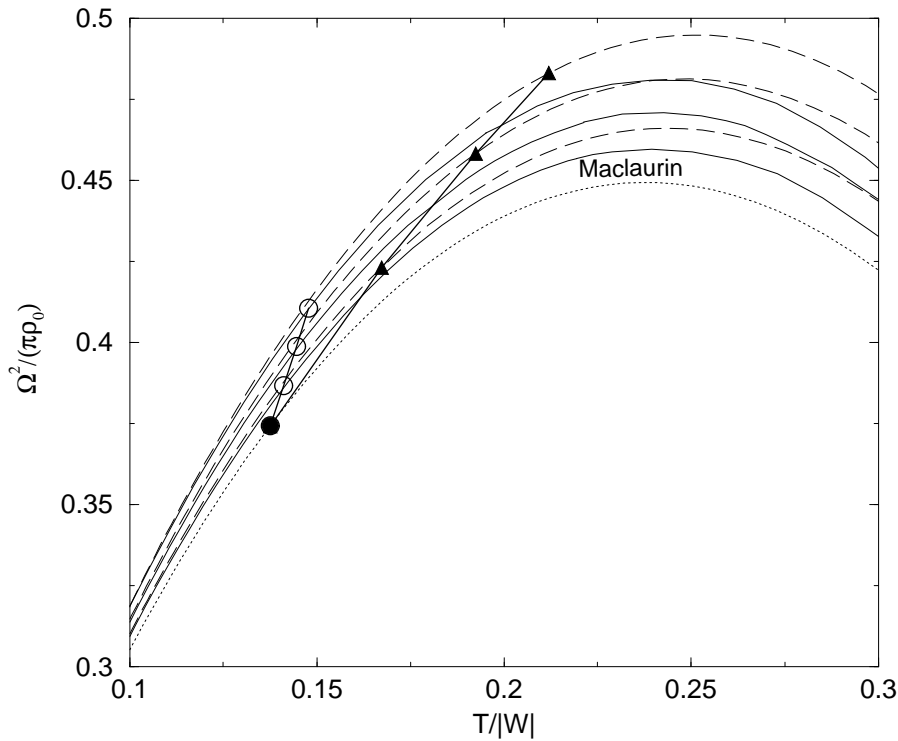}
\caption{\label{f:omeg_tsw} Square of the angular velocity as a
function of the ratio of the kinetic energy to the absolute value of
the gravitational potential energy for the same equilibrium sequences
as in Fig.~\ref{f:omeg_ecc} (all symbols and lines have the same
meaning as in Fig.~\ref{f:omeg_ecc}). The thick lines with circles and
filled triangles correspond to the secular instability points obtained
by us and by \cite{ShapiZ98} respectively.  }
\end{figure}

\section{Calculations and results} \label{s:cal}

\subsection{Rotating axisymmetric equilibrium configurations}

We present in this section our results about the relativistic analogs of
the axisymmetric Newtonian Maclaurin ellipsoids.
As mentioned in Sec.~\ref{s:num_tech}, the multi-domain
spectral technique with surface-fitted coordinates that we
employ has some trouble when $r_{\rm p}/r_{\rm eq} < 1/3$.
To compute very rapidly rotating configurations, well beyond
the Jacobi-like bar mode instability (which is located
at $r_{\rm p}/r_{\rm eq} \simeq 0.58$ in the Newtonian regime),
we employ a second numerical code for axisymmetric stationary
relativistic stars, namely that of Stergioulas \cite{StergF95}
(see \cite{Sterg98} for a description).
More precisely, the multi-domain spectral code has
been used for calculating rotating stars
with the eccentricity $e$ (or $T/|W|$) lower than $0.92-0.87$
(0.23-0.2) for compaction parameters ranging from 0 to 0.25,
while Stergioulas' code has been used  for
higher values. In the overlapping region around
$e\sim 0.9$, models computed by both codes agree very well in
all computed properties.
  
We have constructed constant baryon mass sequences of uniformly rotating
axisymmetric homogeneous fluid bodies. Each of
the sequences is parametrized by the compaction parameter $M_{\rm s}/R_{\rm s}$,
where $M_{\rm s}$ and $R_{\rm s}$ are the gravitational mass and
circumferential radius of the
spherical (non-rotating) member of the sequence.  To show the role of
relativistic effects we plot in Fig.~\ref{f:omeg_ecc_ours} and
Fig.~\ref{f:omeg_tsw_ours} the square of the angular velocity as a
function of the eccentricity $e$ and $T/|W|$ respectively for several
axisymmetric equilibrium sequences (thin lines) with compaction
parameters $M_{\rm s}/R_{\rm s}$ ranging from 0 (Newtonian Maclaurin sequence) to
0.25.  We find that in general relativity the rotational velocity
along the equilibrium sequence is not much different from that one
derived in the Newtonian limit, although a star of a given
eccentricity has a little bit higher value of the square of the
angular velocity (the biggest difference is near the maximum, and for
$M_{\rm s}/R_{\rm s}=0.25$, it is by $~35\%$ higher than the Newtonian one). The
location of the maximum of $\Omega^2/(\pi \rho_0)$ weakly depends on the
compaction parameter and is only a slightly shifted to lower values of
eccentricity and higher values of $T/|W|$ for highly relativistic
stars.

In Figs.~\ref{f:omeg_ecc} and ~\ref{f:omeg_tsw}
we compare our fully relativistic sequences with
PN equilibrium sequences in the ellipsoidal approximation
\cite{ShapiZ98}, for
compaction parameters $M_{\rm s}/R_{\rm s}=0.025$,
$0.05$ and  $0.075$. Let us recall that, contrary to the
Newtonian case, in general relativity the figure of equilibrium of rotating
incompressible bodies is not an ellipsoid (with respect
to the employed coordinates).
However as it can be seen for low compactness
the ellipsoidal PN models and fully relativistic models are in close
agreement up to the value of eccentricity $\sim 0.9$ and $T/|W| \sim 0.2$.
The discrepancy increases as eccentricity increases. For
$M_{\rm s}/R_{\rm s}=0.075$ and $e=0.9$ the value of
$\Omega^2/(\pi \rho_0)$ is overestimated by $3\%$ using
the ellipsoidal approximation.

For high compaction parameter ($M_{\rm s}/R_{\rm s}>0.1$)
comparison between our fully
relativistic calculations (Figs.~\ref{f:omeg_ecc_ours} and
\ref{f:omeg_tsw_ours}) and PN ones (Fig. 3 in \cite{ShapiZ98}) shows
 that the ellipsoidal approximation does not provide a good description of
 equilibrium axisymmetric configurations. We find a more pronounced
 increase of the angular velocity with compactness at any
 given value of eccentricity. The relative differences are $~10\%$ and
 $~20\%$ for $M_{\rm s}/R_{\rm s} =0.15$ and $0.25$ respectively and
 $e<0.8$ (there are much larger for higher value of $e$).  Moreover
 according to \cite{ShapiZ98} the location of the maximum of
 $\Omega^2/(\pi \rho_0)$ is shifted to higher values of $e$ with
 respect to Newtonian one, while we find the opposite tendency, in
 agreement with \cite{GirolV02}.

\begin{table*}
\caption{\label{t:instab} {\bf Jacobi-like instability points
along relativistic sequences.} The symbols
are as follows (all values are for axisymmetric configurations at
the instability point): $H_c$ is the central enthalpy;
${M}\over{R_{\rm circ}}$ is the proper compaction parameter, where
$R_{\rm circ}$ is circumferential radius; ${M_{\rm s}}\over{R_{\rm s}}$ is the
compaction parameter of nonrotating spherical configurations with the
same rest mass as the marginally stable configuration; $N_c$ is central lapse;
$\Omega^2/(\pi\rho_0)$ is a square of the
angular velocity in the unit $\pi G \rho_0$; $e_{\rm crit}$ and ${\rm
(T/|W|)_{\rm crit}}$ are eccentricity and the ratio of the kinetic
energy to the absolute value of the gravitational potential energy at
the instability points; GRV2 and GRV3 are virial errors.}
\begin{ruledtabular}
\begin{tabular}{ccccccccc}
%
%
$H_c$ & ${M_{\rm s}}\over{R_{\rm s}}$& ${M}\over{R_{\rm circ}}$ & $N_c$ & $\Omega^2/(\pi\rho_0)$
& $e_{\rm crit}$ & ${\rm (T/|W|)_{\rm crit}}$ & GRV2 & GRV3  \\ \hline
---   & 0     & 0      &  1      & 0.37423 & 0.81267 & 0.13753  &  ----  &  3.e-12\\
0.01  & 0.025 & 0.0227 & 0.95976 & 0.38670 & 0.81528 & 0.14119  & -3e-07 & -5.e-05 \\
0.02  & 0.050 & 0.0438 & 0.92172 & 0.39885 & 0.81750 & 0.14464  &  2e-06 &  5e-05 \\
0.04  & 0.094 & 0.0815 & 0.85174 & 0.42222 & 0.82096 & 0.15097  & -3e-07 &  6e-05 \\
0.05  & 0.113 & 0.0984 & 0.81955 & 0.43346 & 0.82231 & 0.15389  & -4e-06 & -2e-05 \\
0.07  & 0.151 & 0.1289 & 0.76022 & 0.45512 & 0.82439 & 0.15930  &  8e-07 &  9e-05 \\
0.09  & 0.181 & 0.1556 & 0.70691 & 0.47570 & 0.82584 & 0.16422  & -2e-05 &  -5e-05 \\
0.11  & 0.205 & 0.1792 & 0.65885 & 0.49528 & 0.82684 & 0.16871  &  9e-06 &  2e-05 \\
0.12  & 0.219 & 0.1899 & 0.63658 & 0.50474 & 0.82721 & 0.17083  & -1e-06 &  -7e-05 \\
0.13  & 0.230 & 0.2000 & 0.61537 & 0.51403 & 0.82755 & 0.17290  & -1e-05 & -1e-4 \\
0.15  & 0.250 & 0.2187 & 0.57593 & 0.53192 & 0.82795 & 0.17674  &  2e-05 & 3e-05 \\
0.16  & 0.258 & 0.2272 & 0.55757 & 0.54057 & 0.82809 & 0.17859  &  1e-05 & -6e-05 \\
0.18  & 0.275 & 0.2430 & 0.52322 & 0.55748 & 0.82838 & 0.18220  & -3e-05 & -2e-4 \\
\end{tabular}
\end{ruledtabular}
\end{table*}

\subsection{Secular bar mode instability} \label{s:resu_inst}

\subsubsection{Our results}

Some important quantities at the viscosity driven instability points
for different values of compaction parameter are reported in
Table~\ref{t:instab}.
We are using two kinds of compaction parameters: $M_{\rm s}/R_{\rm s}$
already defined and the {\em proper
compaction parameter} $M/R_{\rm circ}$, where $R_{\rm circ}$ is the
circumferential radius, i.e. the length of the equator (as given by
the metric) divided by $2\pi $ of the actual configuration
(unlike $R_{\rm s}$ which is the circumferential radius of the
non rotating star having the same baryon mass).

Our results are shown as thick lines in Figs.~\ref{f:omeg_ecc_ours},
\ref{f:omeg_tsw_ours} and \ref{f:ecc_comp}. According to our analysis
the critical value of the eccentricity very weakly depends on the
compaction parameter and for $M_{\rm s}/R_{\rm s}=0.275$ is by only
$2\%$ larger than Newtonian value of the onset of the secular bar mode
instability.  The critical value of the ratio of the kinetic energy to
the absolute value of the gravitational potential energy $(T/|W|)_{\rm
crit}$ for compaction parameter as high as 0.275 is by $~30\%$
(Table~\ref{t:instab}) higher than the Newtonian value.
The dependence of $(T/|W|)_{\rm crit}$ on the compactness
can be very well approximated by the function
\begin{equation}
(T/|W|)_{\rm crit} \simeq (T/|W|)_{\rm crit,Newt}
	+ \left\{ \begin{array}{ll}
	0.148 \, x(x+1) & \quad\mbox{for} \quad x := M/R_{\rm circ} \\
	0.126 \, x(x+1) & \quad\mbox{for} \quad x := M_{\rm s}/R_{\rm s}
	\end{array} \right. 
\end{equation}

\subsubsection{Comparison with PN calculations}

The comparison between our relativistic calculations of the viscosity
driven instability and the corresponding two different PN calculations
derived by \cite{ShapiZ98} and by \cite{GirolV02} is shown in Fig~\ref{f:omeg_ecc},
\ref{f:omeg_tsw} and \ref{f:tsw_comp}.
Shapiro and Zane \cite{ShapiZ98} employ an energy variational principle
to determine equilibrium shape and stability of homogeneous triaxial
ellipsoids.  The method used by them is valid for arbitrary rotation
rate, but only for constant density bodies. Di Girolamo and Vietri
 \cite{GirolV02}
determined the value of the eccentricity and $\Omega^2/(\pi \rho_0)$ at
the instability onset point using Landau's theory of second-order
phase transitions.  This method is the extension to PN regime
of that used by Bertin and Radicati \cite{BertiR76} for
the Newtonian treatment of bar mode instability and valid for
any equation of state.  Considering the
dependence of $\Omega^2/(\pi \rho_0)$ with respect to the eccentricity
(Fig.~\ref{f:omeg_ecc}) we found good agreement between our results and
the PN ones of \cite{GirolV02}.  As can be seen, both calculations show
much weaker influence of relativistic effects on location of the
instability onset point than Shapiro \& Zane PN calculations.

Figure~\ref{f:tsw_comp} shows the dependence of the critical
eccentricity and the critical value of $T/|W|$ on the compaction
parameter $M_{\rm s}/R_{\rm s}$. The Authors of \cite{GirolV02} use
a proper ``conformal compaction parameter'' $M_{\rm c}/R_{\rm
c}:=4\pi\rho_0/3R_{\rm c}^2$, where $R_{\rm c}= (a_1 a_2 a_3)^{1/3}$,
$a_1$, $a_2$, $a_3$ being the ellipsoid semiaxes. In
order to make a comparison between our relativistic calculations and
those of \cite{GirolV02} we find the $M_{\rm s}/R_{\rm s}$ corresponding to their
$M_{\rm c}/R_{\rm c}$ using the formula $M_{\rm c}/R_{\rm c}=0.25
M_{\rm s}/R_{\rm s}(1 - M_{\rm s}/R_{\rm s}+(1-2M_{\rm s}/R_{\rm
s}))^2$ (see \cite{LightPPT79}, expression [4], p. 422). This formula
is valid in the spherical limit, but the difference between the proper
``conformal compaction parameter'' and the ``spherical conformal compaction
parameter'' is at most 3 \%.

According to our analysis the critical value of the eccentricity depends very
weakly on the compaction parameter (solid line); this is in
agreement with the PN calculations by Di Girolamo \& Vietri \cite{GirolV02}.
The relative differences between our and their PN
calculations are less than $1\%$.  A much stronger weakening of the
bar mode instability by general relativity is suggested by the
PN study of Shapiro \& Zane \cite{ShapiZ98}, who find that $e_{\rm crit}$
(dashed line in the left panel of Fig.~\ref{f:tsw_comp}) could be
as large as 0.94 at $M_{\rm s}/R_{\rm s}= 0.25$.  This discrepancy may be ascribed
mainly to the ellipsoidal approximation for the deformation and
equilibrium shape.  We refer to the article \cite{GirolV02} for a
detailed explanation of discrepancies between the two PN calculations of instability points.

The right panel of Fig.~\ref{f:tsw_comp} presents the comparison
between \cite{ShapiZ98} and our calculation of the critical $T/|W|$ as
a function of the compaction parameter $M_{\rm s}/R_{\rm s}$. We found
that for the compaction parameter as high as 0.05; 0.15; 0.25
$(T/|W|)_{\rm crit}$ is only by $5\%;$ $16\%; 28\%$ higher than
Newtonian value, while according to Ref.~\cite{ShapiZ98}, the increase
is $40\%; 79\%; 88 \% $ respectively.

\begin{figure}
\includegraphics[height=6cm]{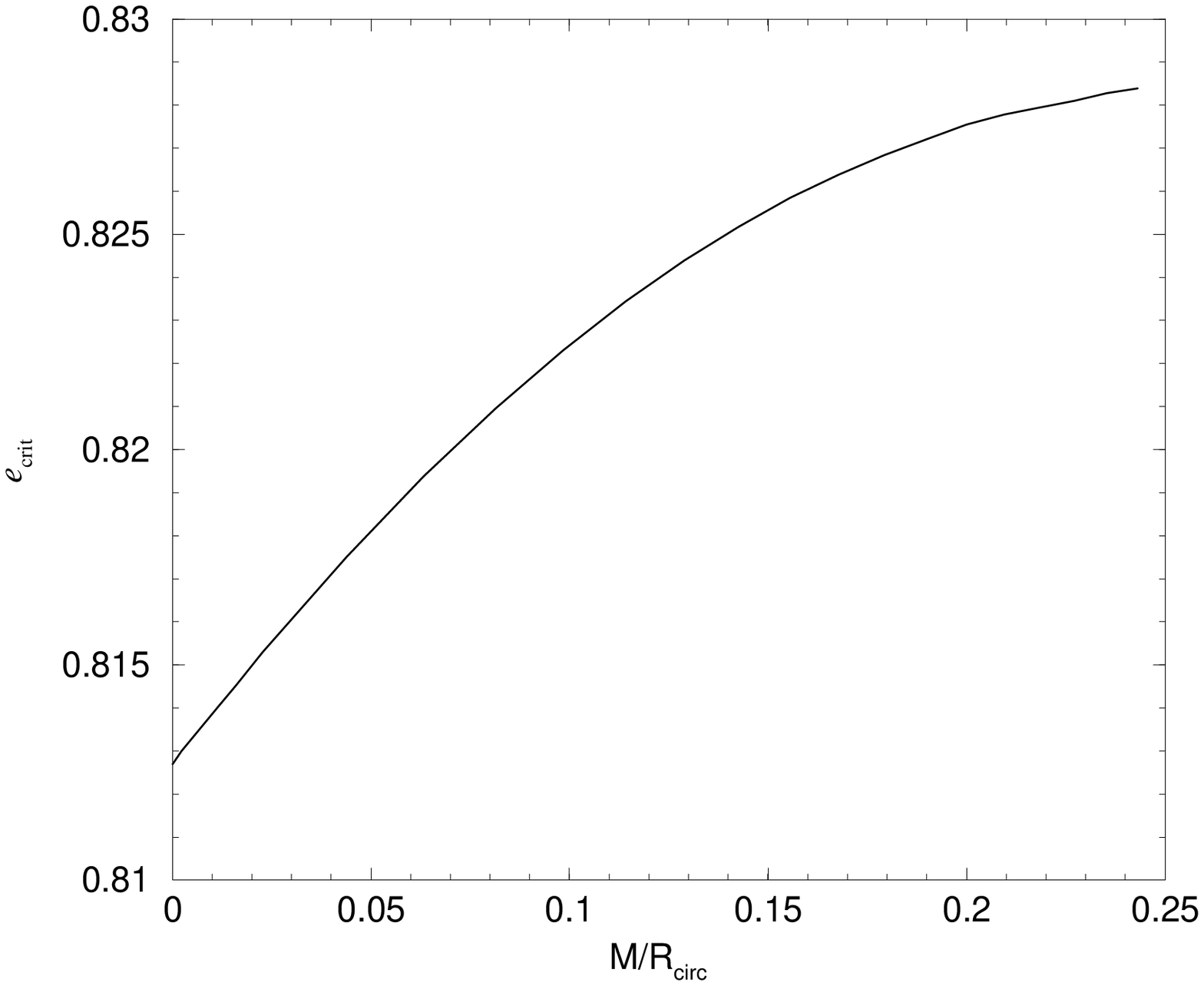} \qquad
\includegraphics[height=6cm]{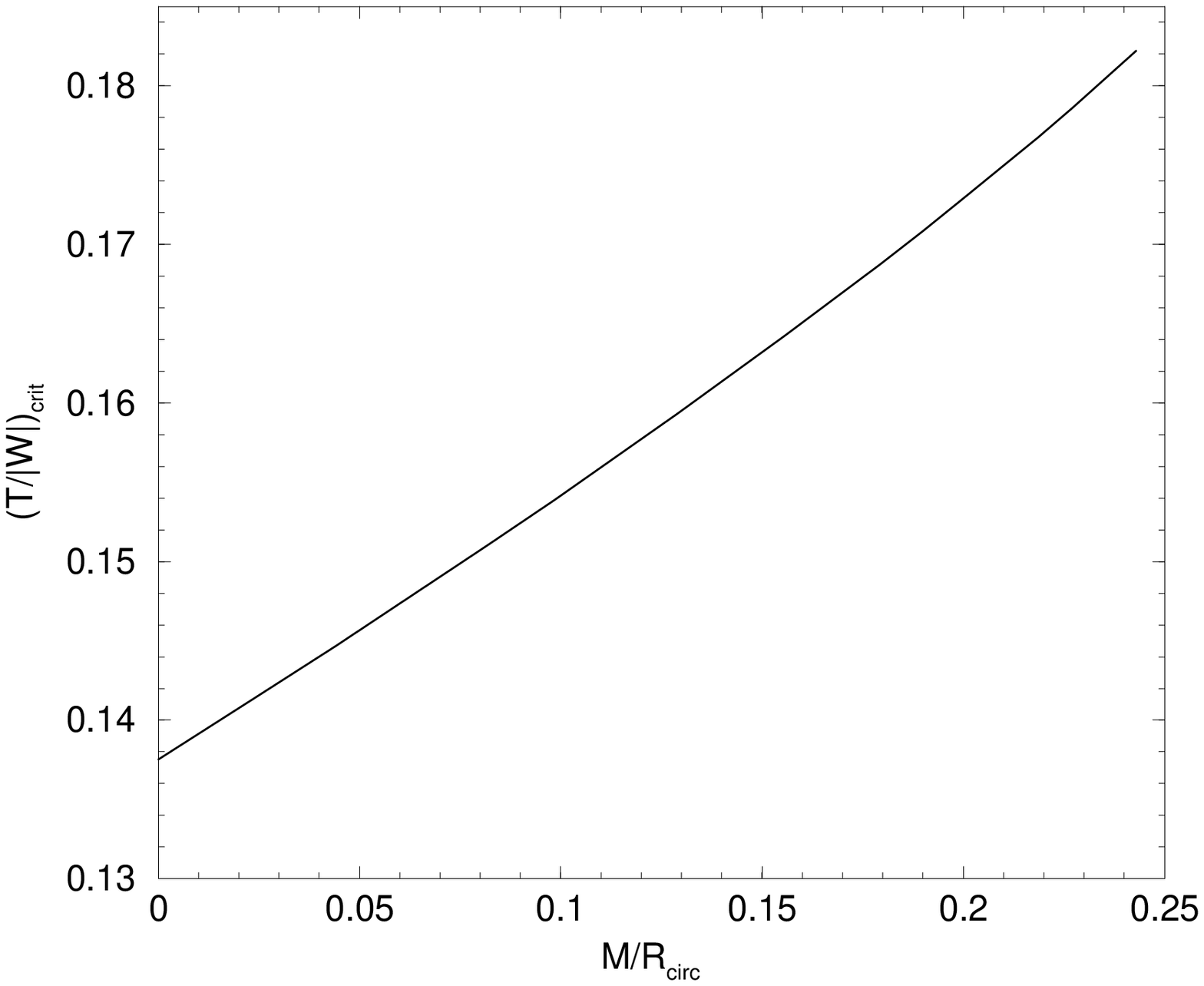}
\caption{\label{f:ecc_comp} Eccentricity (left) and ratio of the kinetic
energy to the absolute value of the gravitational potential energy (right)
at the onset of the secular bar mode instability,
versus the compaction parameter $M/R_{\rm circ}$.}
\end{figure}

\begin{figure}
\includegraphics[height=6cm]{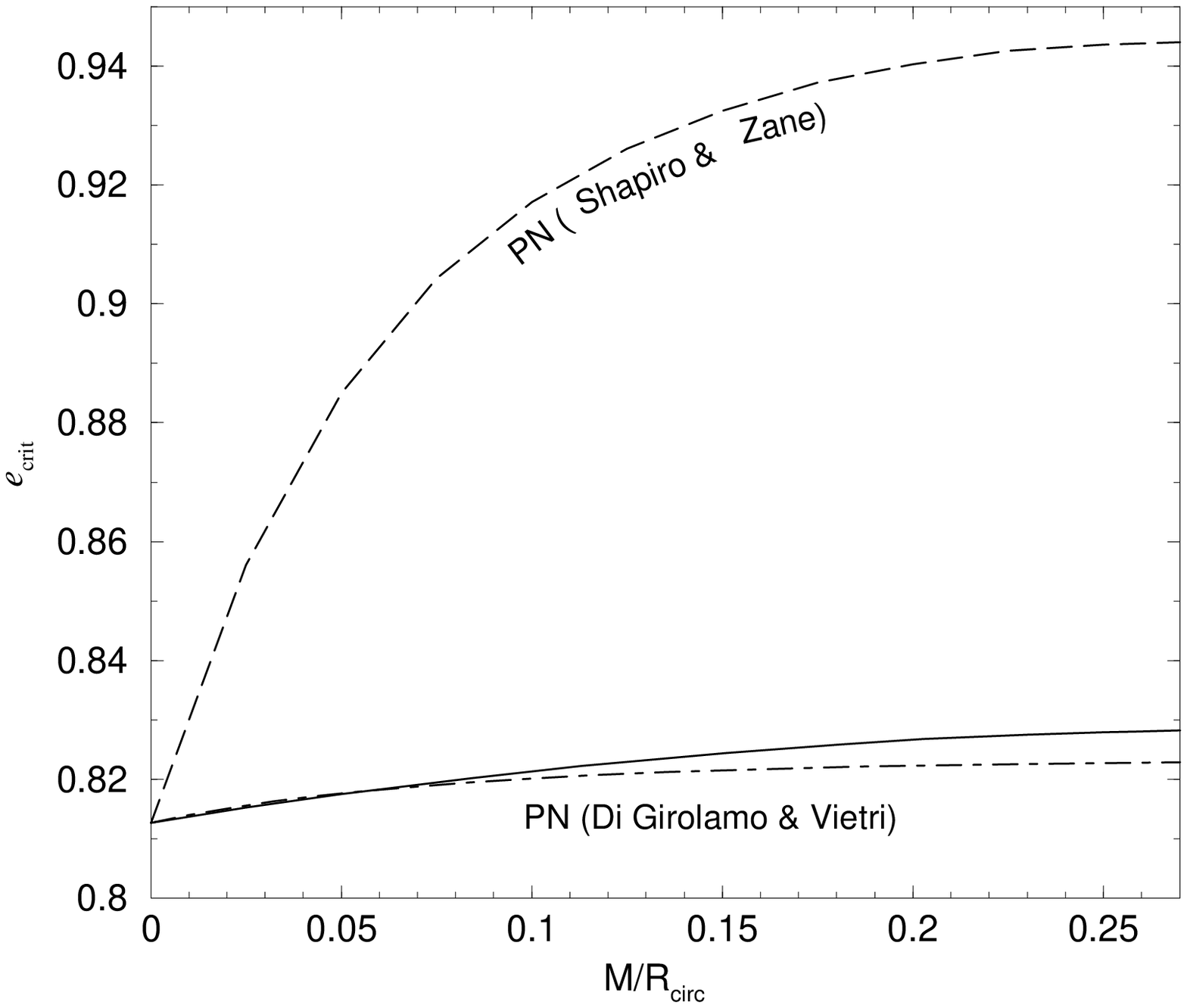}   \qquad
\includegraphics[height=6cm]{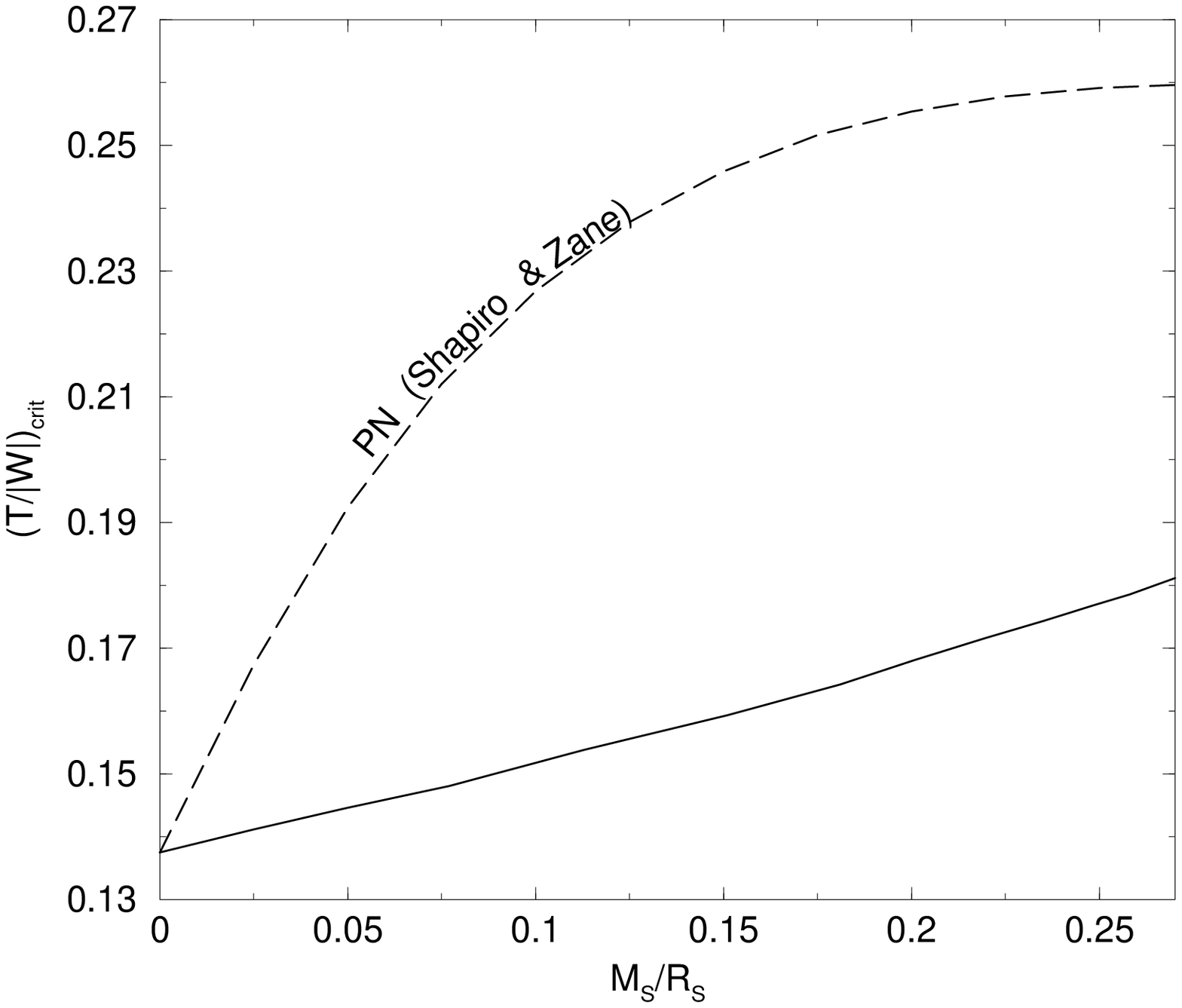}
\caption{\label{f:tsw_comp}
Eccentricity (left) and ratio of the kinetic
energy to the absolute value of the gravitational potential energy (right)
at the onset of the secular bar mode instability,
versus the compaction parameter $M_{\rm s}/R_{\rm s}$.
Our results are denoted by the solid line.  The dashed line and the
dash-dotted line correspond to the PN calculations by \cite{ShapiZ98}
and \cite{GirolV02} respectively.}
\end{figure}

According to our study relativistic effects weaken the Jacobi-like
bar mode instability (solid line in Figs.~\ref{f:omeg_ecc_ours},
\ref{f:omeg_tsw_ours}, \ref{f:ecc_comp} and \ref{f:tsw_comp}),
but the stabilizing effect is not very strong. This
is in agreement with the PN calculations \cite{GirolV02}.



%


\section{Summary} \label{s:summ}

Triaxial instabilities of rotating compact stars can play an important
role as emission mechanisms of gravitational waves in the frequency
range of the forthcoming interferometric detectors.  A rapidly
rotating neutron star can spontaneously break its axial symmetry if
the ratio of the rotational kinetic energy to the absolute value of
the gravitational potential energy $T/|W|$ exceeds some critical
value. We have investigated the effects of general relativity upon the
nonaxisymmetric viscosity-driven bar mode instability of
incompressible, uniformly rotating stars. This is the relativistic
analog of the Newtonian Maclaurin-Jacobi bifurcation point.

Our method of finding the instability point is similar to that
used in Ref.~\cite{BonazFG98} for compressible fluid stars.
The main improvement with respect to this work regards the
numerical technique. We have indeed introduced
surface-fitted coordinates, which enable us to
treat the strongly discontinuous density profile at the surface of
incompressible bodies. This avoids any Gibbs-like phenomenon and results in
a very high precision, as demonstrated by comparison with the
analytical result for the Newtonian Maclaurin-Jacobi bifurcation point,
that our code has retrieved with a relative error of $10^{-6}$.

According to our results, general relativity weaken the Jacobi-like
bar mode instability: the values of $T/W$, eccentricity
and $\Omega^2/(\pi \rho_0)$ increase at the onset of instability above
the Newtonian values. This general tendency is in
agreement with PN analytical results \cite{ShapiZ98,GirolV02} for rigidly
rotating incompressible bodies and with the numerical calculations of
\cite{BonazFG98} for relativistic polytropes. However we found that
the stabilizing effect of general relativity is much weaker than
that obtained in the PN treatment of \cite{ShapiZ98}.  The critical
value of the ratio of the kinetic energy to the absolute value of the
gravitational potential energy $(T/|W|)_{\rm crit}$ for a compaction
parameter as high as 0.275 is only $~30\%$ larger than the Newtonian value,
whereas it has been found $90\%$ larger by \cite{ShapiZ98}.

 According to our analysis the critical value of the eccentricity very
weakly depends on the compaction parameter and for a compaction
parameter as high as 0.275 is only $2\%$ ($16\%$ according to
~\cite{ShapiZ98}) larger than the Newtonian value of the onset of the
secular bar mode instability.  Regarding the dependence of $\Omega^2/(\pi
\rho_0)(e_{\rm crit})$ and $e_{\rm crit}$ with respect to compactness,
we found very good agreement between our result and the recent PN ones
of \cite{GirolV02}, the  relative differences being lower than $1\%$.


\begin{acknowledgments}
 We are grateful to Nick Stergioulas and Tristano Di Girolamo for helpful
 discussions and to Brandon Carter for reading the manuscript.
 We also thank Stuart Shapiro and Silvia Zane for providing tables of their
 results and Tristano Di Girolamo and Mario Vietri for providing
 their results prior to publication. This work has been funded by the
 following grants: KBN grants 5P03D01721; the Greek-Polish Joint
 Research and Technology Program EPAN-M.43/2013555 and the EU
 Program ``Improving the Human Research Potential and the
 Socio-Economic Knowledge Base'' (Research Training Network Contract
 HPRN-CT-2000-00137).
\end{acknowledgments}

\end{document}